\documentclass[letter]{aa}
\usepackage{graphicx}
\usepackage{amsmath}	
\usepackage{amssymb}	
\usepackage{relsize}
\usepackage{txfonts}
\usepackage[colorlinks=true,linkcolor=blue,citecolor=blue,filecolor=blue,urlcolor=blue]{hyperref}

\begin{document} 

   \title{ZF-UDS-7329: A relic galaxy in the early Universe}
   \subtitle{}

   \author{Eduardo A. Hartmann \inst{1,2}, Ignacio Mart\'{\i}n-Navarro \inst{1,2}, Marc Huertas-Company \inst{1,2}, 
           João P. V. Benedetti \inst{1,2}, Patricia Iglesias-Navarro \inst{1,2}, Alexandre Vazdekis \inst{1,2}, 
           Mireia Montes \inst{3}
           }
   \institute{
    Instituto de Astrof\'{\i}sica de Canarias, c/ V\'{\i}a L\'actea s/n, E38205 - La Laguna, Tenerife, Spain\\
    \email{eduardo.hartmann@iac.es}
    \and
    Departamento de Astrof\'isica, Universidad de La Laguna, E-38205 La Laguna, Tenerife, Spain
    \and 
    Institute of Space Sciences (ICE, CSIC), Campus UAB, Carrer de Can Magrans, s/n, 08193 Barcelona
             }
             

   \date{Accepted 2025 January 7. Received 2025 November 28.}
   
   \titlerunning{ZF-UDS-7329: A relic galaxy in the early Universe}
\authorrunning{Hartmann, E. A.}

  \abstract
   {
    The formation time scales of quiescent galaxies can be estimated in two different ways: by their star formation history and by their chemistry. Previously, the methods yielded conflicting results, especially when considering $\alpha$-enhanced objects. This is primarily due to the time resolution limitations of very old stellar populations, which prevent accurately constraining their star formation histories. We analysed the JWST observations of the extremely massive galaxy ZF-UDS-7329 at z$\sim$3.2 and show that the higher time resolution necessary to match the chemical formation time scales using stellar population synthesis can be achieved by studying galaxies at high redshift. We compare the massive galaxy to the well-known relic galaxy NGC~1277, arguing that ZF-UDS-7329 is an early Universe example of the cores of present-day massive elliptical galaxies or, if left untouched, a relic galaxy.
   }


   \maketitle
%

\section{Introduction} \label{sec:intro}

New studies and surveys making use of the James Webb Space Telescope (JWST) have discovered a large population of very massive galaxies with a significant quiescent fraction that appear to have formed their stars very early in the history of the Universe \citep{Valentino2023,Carnall2023,Nanayakkara2024,Weibel2024}. This poses new challenges to the prevailing theory of galaxy formation within the $\Lambda$CDM cosmological model, especially concerning their early and fast formation and subsequent quenching.

In this letter, we focus on the extremely massive (M$_*$$\sim$10$^{11}$\,M$_{\odot}$) galaxy ZF-UDS-7329, which is quiescent at z$\sim$3.2 \citep{Glazebrook2024,Carnall2024,Turner2024}. This galaxy is reported to have formed the majority of its stars in a fast burst lasting between 200 and 400\,Myr before the first billion years of the Universe. \citet{Glazebrook2024} point out that due to this rapid and massive formation, this object is in tension with the current estimations of dark matter halo sizes at this early epoch. This can be somewhat mitigated if different star formation history (SFH) priors are examined and major mergers are considered \citep{Turner2024}. \citet{Carnall2024} have also performed an analysis of this galaxy and indicate similar time scales for formation and quenching. They used the extreme value statistics approach \citep{Lovell2023} to compare the SFH of this galaxy with the most massive objects expected in the survey area. They find that ZF-UDS-7329 is more massive than expected when a fiducial model is considered; however, a more extreme formation efficiency  or significant merger events \citep[e.g. Figure 8 from][]{Turner2024} can lighten or eliminate the tension with the dark matter halo mass function.

Interestingly, these massive quiescent galaxies at high redshift show remarkably similar properties to well-known local objects such as massive relic galaxies. These relics are compact, with an effective radius of a few kiloparsecs \citep{Trujillo2014,Ferre-Mateu2017,Yildirim2017}, and very massive (M$_* \sim 10^{11} - 10^{12}$\,M$_{\odot}$), with a thick-disc morphology \citep{Yildirim2017} and old stellar populations. Briefly, when considering the paradigm of hierarchical formation and assembly \citep{Oser2010}, a small percentage of galaxies would suffer no significant merger events throughout their lives and thus conserve their original formation characteristics \citep{Quilis2013}. A number of them have been found in the local Universe \citep{Spiniello2021,Spiniello2024} and up to z$\sim$1.0 \citep{Lisiecki2023}, with NGC~1277 being the prototypical example \citep{Trujillo2014}. In general, these galaxies are compact, harbour super massive black holes \citep{Scharwchter2016}, have fast rotation curves and disc-like morphologies, and exhibit relatively low dark matter-to-baryon fractions \citep{Yildirim2017,Comeron2023}. One of the issues found when exploring the SFH of these objects using stellar population synthesis is that old stellar populations change their spectra very gradually \citep{Spiniello2021} due to the slow evolution of low-mass stars. Therefore, the best simple stellar population (SSP) models available are not capable of discerning formation episodes shorter than $\sim$0.5-1~Gyr in ideal conditions when considering $\sim$13~Gyr old populations. This imposes a limit to the derived time scale of star formation on such galaxies.

\begin{figure*}[h]
\centering
\includegraphics[width=16.5cm]{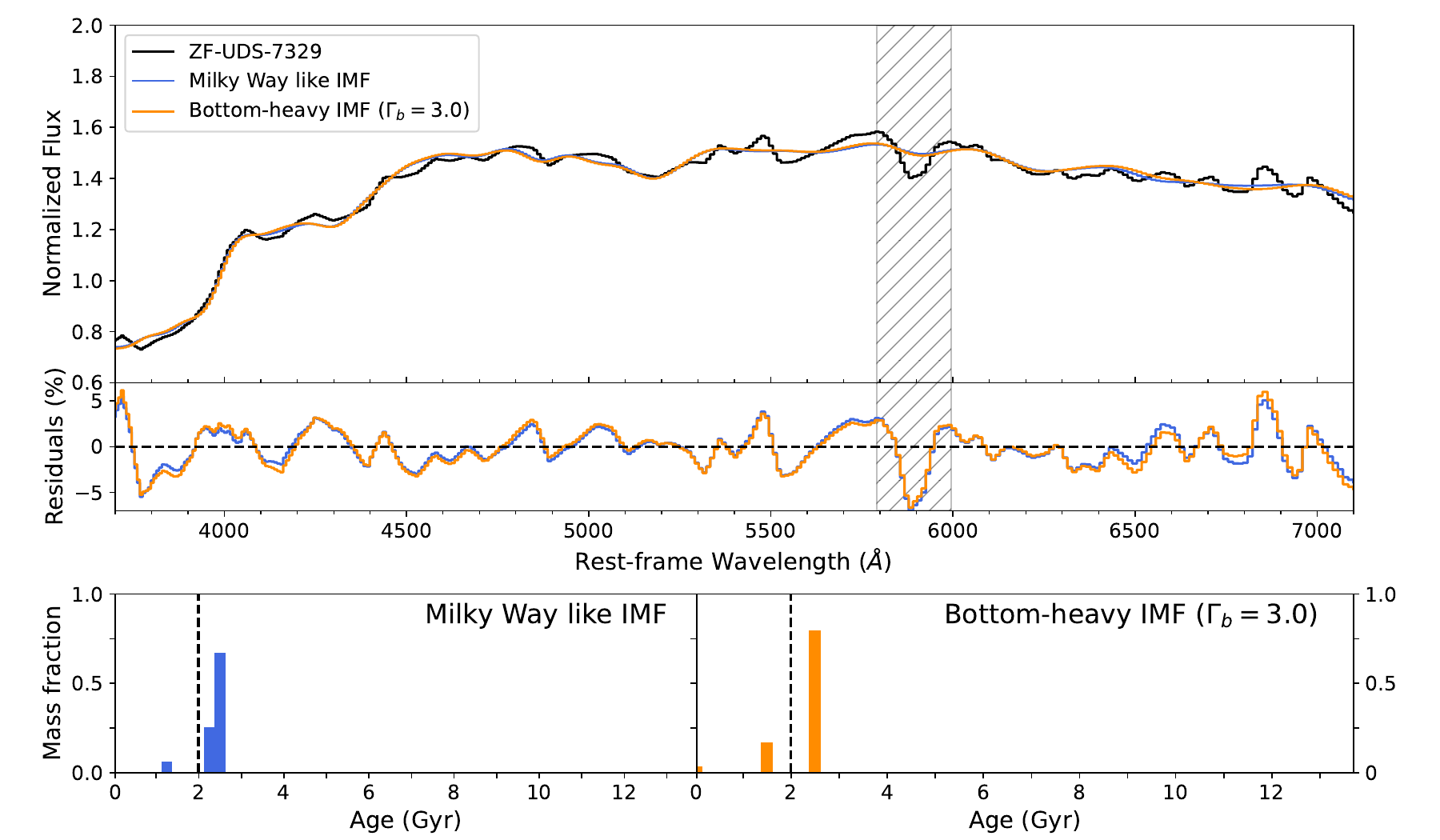}
\caption{Best-fitting models for ZF-UDS-7329. The top panel shows the original PRISM spectra in black and the two best-fit models from pPXF. In blue is the Milky Way-like Kroupa set of SSPs and in orange is the bottom-heavy IMF $\Gamma_{\text{b}}=3.0$ (this colour scheme is throughout the letter). The middle panel shows the residuals between the models and observations, where the shaded region is the sodium doublet mask. The bottom panels show the mass fractions as a function of age.}
\label{fig:ppxf_fit}
\end{figure*}

A different approach to obtaining the star formation time scale of such massive galaxies that can help mitigate the issue mentioned is to use their stellar population's chemistry \citep{Worthey1992,Thomas2005,delaRosa2011,Martin-Navarro2016}. A well-known property of massive quiescent galaxies is the correlation of the [Mg/Fe] abundance with stellar mass and velocity dispersion \citep{Worthey2003,Pernet2024}. The prevailing explanation for this is short and intense star formation episodes where Type II supernovae enrich the interstellar medium with $\alpha$-elements. The star formation is then rapidly quenched before a significant amount of Type Ia supernovae can replenish the supply of Fe-peak elements, causing an enhancement in [$\alpha$/Fe]. Based on this, measured values of [Mg/Fe] can be translated into chemical star formation time scales \citep{Thomas2005}. However, when comparing the SFH of local massive relic galaxies and their chemical formation time scales, discrepancies arise. The formation time scales obtained based on SFHs are systematically longer than those expected from the measured [Mg/Fe], in particular for the most massive and rapidly formed galaxies in the local Universe \citep{McDermid2015}.

Another aspect to consider is that the slope of the initial mass function (IMF) of relic galaxies and the cores of massive ellipticals \citep{Martin-Navarro2015a,Martin-Navarro2015b,vanDokkum2017,Maksymowicz-Maciata2024} have been found to be steeper than the standard Milky Way IMF \citep{Kroupa2001}. \citet{Martin-Navarro2016} applied these findings to the estimation of the chemical formation time scale and showed that it predicts even shorter episodes of star formation in these galaxies, widening the gap between chemical and SFH time scales. Nonetheless, a simple bottom-heavy IMF is inconsistent with the observed chemistry of these objects unless a time-varying IMF is invoked \citep{Vazdekis1997,Weidner2013,Ferreras2015}. While some works have argued for a top-heavy IMF in high-$z$ star forming systems \citep{Sneppen2022,Cameron2024}, it is worth noting that such massive stars do not contribute to the stellar continuum of quiescent galaxies older than $\sim$1~Gyr.

In this letter, we explore the connection between the high-z massive quiescent galaxy ZF-UDS-7329 and local massive relics. We derive the SFH of ZF-UDS-7329, unconstrained by its redshift, showing that it formed very early and rapidly. We passively age the galaxy's spectrum and show its remarkable similarity to a local massive relic. We also compare the SFH of ZF-UDS-7329 to a sample of nearby quiescent galaxies and show that the increased time resolution in the early Universe can help address the discrepancies between the different ways of estimating the star formation time scales. The combination of the high spatial resolution and detailed studies of local objects with the time resolution advantages offered by high-z galaxies is crucial to forming a complete picture of galaxy formation.
We adopt the flat $\Lambda$CDM cosmology of \citet{Planck2020} with $\Omega_{\text{M}}=0.310$ and H$_{0}=67.7$~km~s$^{-1}$~Mpc$^{-1}$.

\section{Data and methods} \label{sec:pPX}

ZF-UDS-7329 was observed using the PRISM disperser on the Near Infrared Spectrograph (NIRSpec) instrument aboard JWST and spans a wavelength range of 0.6-5.0\,$\mu$m (1,420-12,590\,$\mathring{\text{A}}$ rest frame) with R$\sim$100. The spectrum was kindly provided by Dr. Glazebrook. Further details on the modes used and the reduction process are described in \citet{Glazebrook2024}.

We used the MILES single stellar population models \citep{Vazdekis2010} that span a wavelength range from 3,540 to 7,410\,$\mathring{\text{A}}$ and a resolution of 2.51\,$\mathring{\text{A}}$. The full set of models was generated with BaSTI isochrones, with the 53 ages varying from 0.03\,Gyr to 14\,Gyr and ten metallicities [M/H] from -1.79 to +0.26. Two sets of models were used with different IMF slopes: one with a canonical Milky Way-like Kroupa IMF \citep{Kroupa2001} and a second one with the measured IMF slope of NGC1277 \citep{Martin-Navarro2015b}, that is, a low-tapered bimodal IMF with the logarithmic slope of the upper segment ($>0.6$~M$_{\odot}$) $\Gamma_{\text{b}}=3.0$.

\begin{figure*}[h]
\includegraphics[width=16.5cm]{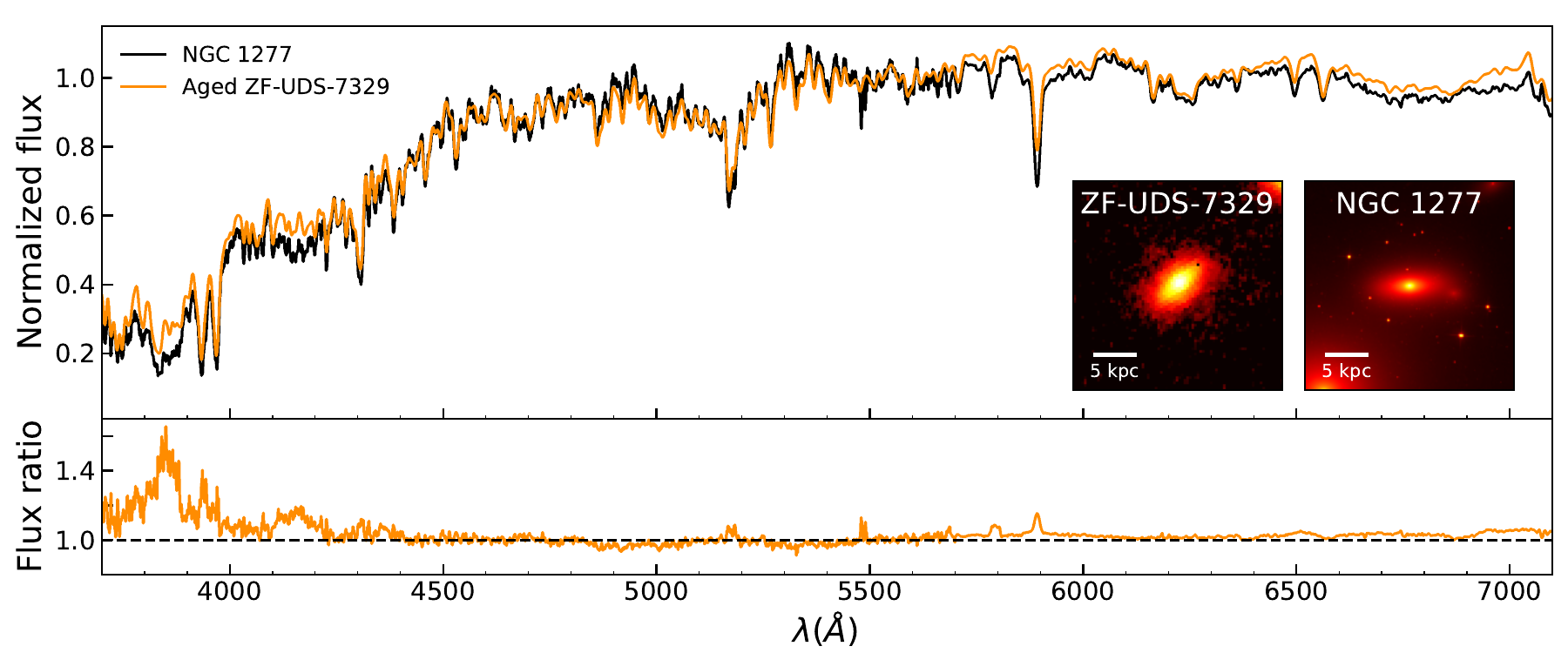}
\caption{Comparison of the aged version of ZF-UDS-7329 (z=3.2) and the massive relic galaxy NGC~1277 (z=0). The inset images of ZF-UDS-7329 and NGC~1277 are from the F444W (JWST) and F814W (Hubble Space Telescope) filters, respectively. The remarkable similarity between the two spectra is evident, and the major differences appear in important absorption features, such as the CN bands, Mg, and Na doublet lines. This is likely due to the particular chemistry of NGC~1277 and the continuum systematics on its observed spectrum.}
\label{fig:comp_bimodal}
\end{figure*}

We measured the mean age, metallicity, and SFH of ZF-UDS-7329 combining the MILES models with the penalised pixel-fitting (pPXF) method from \citet{Cappellari2004}, updated in \citet{Cappellari2023}. Briefly, it is an inversion algorithm that when given an SSP basis attempts to find the best linear combination to minimise the $\chi^{2}$ between the model and observation. Due to the high instrumental dispersion, no kinematic information could be reliably obtained. We included a low-order (n=3) multiplicative Legendre polynomial to correct for any instrumental bias and masked the sodium doublet line between 5,790 and 5,995\,$\mathring{\text{A}}$. The depth of this line is likely due to the presence of neutral gas in the galaxy \citep[e.g.][]{Belli2024}. The SSPs and the spectrum of ZF-UDS-7329 were all normalised so that the weights obtained using pPXF are the mass percentages assigned to each stellar population.

\section{Analyses}

\subsection{Full spectral fitting}
\label{sec:ppxf}

We began our analyses by running pPXF using the Kroupa IMF set of models. The results are shown in Figure~\ref{fig:ppxf_fit}. The top panel shows the original PRISM spectra for ZF-UDS-7329 in black and the pPXF best-fitting solution in blue. The residual is shown in the middle panel. The bottom-left panel shows the mass weights of each population as a function of age. We obtained a mean mass-weighted age of 2.33\,Gyr and [M/H]=0.23 (1.89~Gyr and [M/H]=0.06 if light-weighted). The main mass bin has 2.5\,Gyr and [M/H]=0.26, with the second bin being 2.25\,Gyr old and more metal poor ([M/H]=0.06). A contribution from a population 1\,Gyr younger is also present with a fraction of $\sim$5\%. It is important to note that here complete freedom is given to pPXF to combine models from 0.03\,Gyr up to 14\,Gyr, and the resulting mass fractions are unconstrained by any information from the galaxy's redshift. Nevertheless, the oldest population recovered with pPXF is remarkably consistent with the age of the Universe at that redshift (vertical dashed line in Fig. \ref{fig:ppxf_fit}), highlighting the consistency between $\Lambda$CDM cosmology and stellar population model predictions. To assess the stability of the solution and the uncertainties in the stellar populations properties, we performed 10,000 Monte Carlo simulations, perturbing the spectra within the residuals of the best-fitting result using random Gaussian noise and re-running the fit. The results are very stable, with the weights varying less than 3\% and no change in the age or metallicity bins. Varying the level of regularisation within a reasonable range (from 0.001 - 0.1) does not significantly alter the solution either. Higher regularisation values drive the solution towards unphysical ages and metallicities, as expected since regularisation effectively acts as an informative prior.

Motivated by the fact that nearby massive relic galaxies systematically exhibit a relative excess of low-mass stars, in other words a bottom-heavy IMF \citep{Martin-Navarro2015b,Maksymowicz-Maciata2024}, we performed a second fit using the bottom-heavy IMF with a slope of $\Gamma_{\text{b}}=3.0$. The results are shown in orange in Figure~\ref{fig:ppxf_fit} with a slightly more extended SFH when compared to the Milky Way-like IMF \citep[similar to the effect shown in][]{Ferre-Mateu2013}. We obtained a mass-weighted mean age of 2.24~Gyr and [M/H]=0.15 (1.99~Gyr and [M/H]=0.1 if light-weighted). The majority of the mass fraction is present in the 2.5\,Gyr bin (similar to the previous result), and there is a secondary peak at 1.5~Gyr, with both components having [M/H]=0.15. In this case, a small contribution from the youngest SSP (0.03~Gyr) is present with a weight smaller than 3\%.

Two redshift values are present in the literature for ZF-UDS-7329. In this work, we use the value obtained by \citet{Glazebrook2024} of 3.205$\pm$0.005. The age of the Universe at the time of observation is 1.99\,Gyr, and it is represented by the vertical dashed line in the bottom panels of Fig. 1 (the age difference with the redshift value from \citet{Carnall2024} is less than 0.08\,Gyr). This places the oldest SSP found in both cases 0.5\,Gyr before the beginning of the Universe. Tracing the oldest stellar ages in high-z galaxies can place independent constraints on the cosmology, particularly in the era of JWST observations. Nonetheless, it is important to keep in mind that several model systematics can affect the absolute value of the recovered ages \citep[see e.g.][for further discussion]{Vazdekis2010}, and a 0.5\,Gyr shift is well inside current model uncertainties.

\subsection{A 2~Gr old massive relic galaxy}
\label{sec:ageing}

Visually ZF-UDS-7329 is very similar to local massive relic galaxies such as NGC~1277, as can be seen in the insets of Figure~\ref{fig:comp_bimodal}. They are compact and old and show very little to no internal structure. In order to make a more direct comparison between their spectra, we applied a process of passively ageing the best-fitting model spectrum of ZF-UDS-7329. For this we assumed that the galaxy remained quiescent and unperturbed through the rest of its evolution, this being the primary characteristic of the local massive relic galaxies. We took the age of the best-fitting models from pPXF using the bottom-heavy IMF, added the time that has passed since $z=3.2$, and found the equivalent SSPs in today's Universe. Combining the results weighted by mass, we obtained an artificial spectra of ZF-UDS-7329 as it would be observed at $z=0$. Figure~\ref{fig:comp_bimodal} shows the comparison between the aged spectrum of ZF-UDS-7329 (z=3.2) and that of the prototypical massive relic galaxy NGC~1277 from \citet{Trujillo2014}. Both were normalised to the average flux between 5,300 and 5,700~$\mathring{\text{A}}$.

The overall shape of the spectra shown in Fig.~\ref{fig:comp_bimodal} are remarkably similar despite the intrinsic differences and being in very distinct times (high-z versus local Universe). The largest relative difference can be found in the blue region, particularly with NGC~1277 showing deeper absorption in the two CN bands (between $\sim$3,800-3,900\,$\mathring{\text{A}}$ and $\sim$4,100-4,200$\mathring{\text{A}}$). This difference can be attributed to a higher abundance of this molecule, with the overall excess of flux in the blue due in part to the $\alpha$-enhancement of the relic galaxy that is not present in the SSP models used \citep[for further discussion see Section 4.2 in][]{Vazdekis2015}. From 5,700~$\mathring{\text{A}}$ redwards, differences are mainly due to a systematic effect on the NGC~1277 data \citep{Martin-Navarro2015b}. Despite the differences between NGC~1277 and ZF-UDS-7329, their spectra show remarkable similarities, pointing towards a possible direct evolutionary connection between the high-z quiescent galaxies detected with JWST and the local population of massive relic galaxies.

\subsection{Formation time scales: Local versus high-z massive quiescent galaxies}
\label{sec:formation}

\begin{figure}
\centering
\includegraphics[width=8cm]{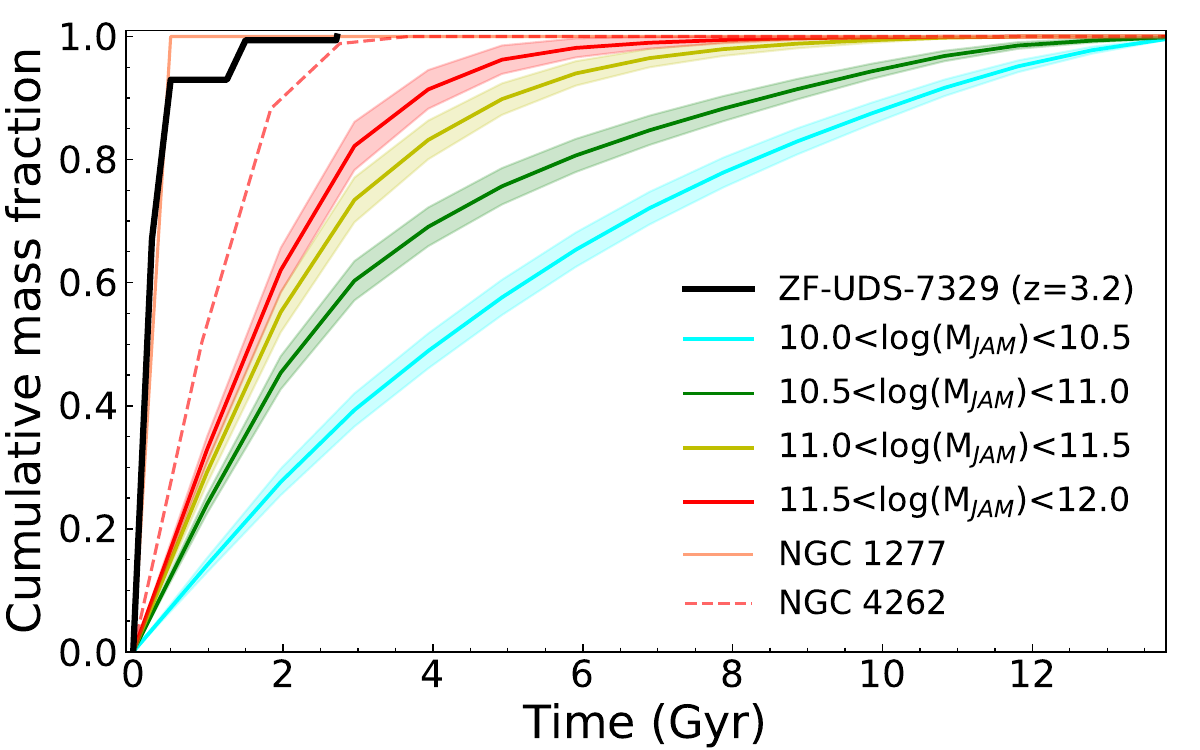}
\caption{Average cumulative mass fraction as a function of time. The ATLAS$^{3\text{D}}$ sample is separated into 0.5\,dex mass bins from 10-10.5\,log~M$_{\text{JAM}}$ in cyan to 11.5-12\,log~M$_{\text{JAM}}$ in red \citep[adapted from Fig. 15 of][]{McDermid2015}. NGC~1277 is shown in orange, and NGC~4262, the fastest growing galaxy in the ATLAS$^{3\text{D}}$ sample, is indicated as a dashed red line. ZF-UDS-7329 is shown in black. Here, we use the result obtained with the Milky Way-like IMF for consistency with the ATLAS$^{3\text{D}}$ sample. The first episode of star formation for every galaxy is set at time zero.}
\label{fig:ATLAS3D}
\end{figure}

It is well established that nearby massive quiescent galaxies formed the majority of their stellar mass very early on in the history of the Universe and have achieved high metallicities \citep{Kauffmann2003,Wake2012}. In Figure~\ref{fig:ATLAS3D}, we show the average cumulative mass fraction as a function of time after the first episode of star formation for the ATLAS$^{3\text{D}}$ sample of nearby early-type galaxies \citep[this is a modified version of Fig. 15 from][]{McDermid2015}. They are separated into 0.5~dex mass bins, starting at 10-10.5\,log~M$_{\text{JAM}}$\footnote{We note that M$_{\text{JAM}}$ is the dynamical mass based on the Jeans anisotropic modelling from \citet{Cappellari2008}} in cyan and moving to green, yellow, and finally red, which represents the 11.5-12\,log~M$_{\text{JAM}}$ bin. Additionally, we show NGC~4262 as a dashed red line (the fastest growing galaxy in ATLAS$^{3\text{D}}$) and the massive relic galaxy NGC~1277. The cumulative mass fraction of ZF-UDS-7329 is shown in solid black.\footnote{For consistency with the ATLAS$^{3\text{D}}$ sample, we show the SFH obtained based on the Milky Way-like IMF}

It is clear that ZF-UDS-7329 grew faster than typical massive galaxies and much faster than any other in the ATLAS$^{3\text{D}}$ sample. On the other hand, the SFH of NGC~1277 is consistent with that measured for ZF-UDS-7329 but limited by the coarse time resolution of old stellar population models. In practice, changes in the spectra of old populations such as those present in nearby massive galaxies are subtle, making it effectively impossible to distinguish between time scales shorter than $\sim$0.5~Gyr. As such, the fact that we can discern that ZF-UDS-7329 formed 65-80\% (depending on the IMF considered) of its mass in less than 250~Myr is a consequence of the increased time resolution that can be achieved by studying the SFH of high-z objects.

\begin{figure}
\centering
\includegraphics[width=8cm]{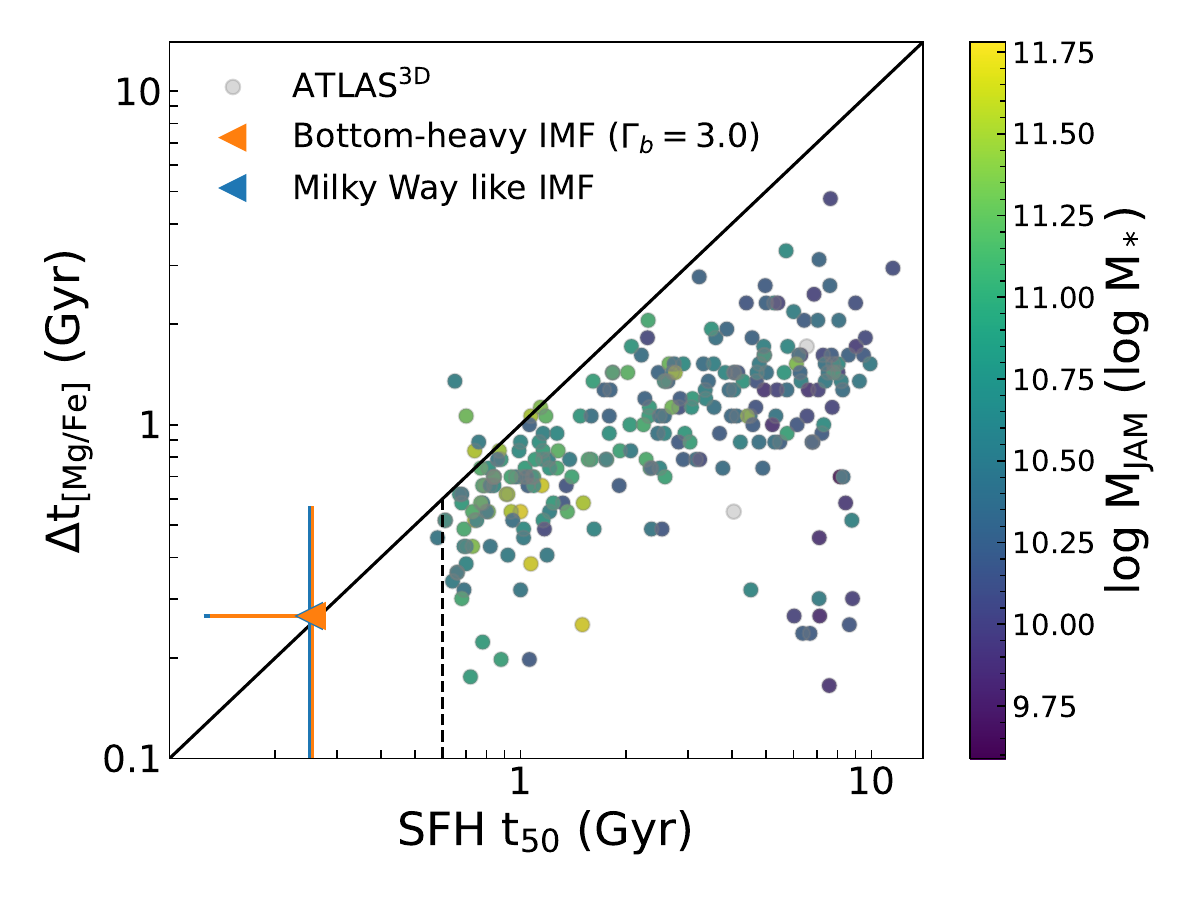}
\caption{Chemical (t$_{[\text{Mg}/\text{Fe}]}$) versus SFH (t$_{50}$) formation time scales. Circles are the ATLAS$^{3\text{D}}$ sample with t$_{[\text{Mg}/\text{Fe}]}$ calculated using the expression from \citet{Thomas2005}, colour-coded according to M$_{\text{JAM}}$. In blue and orange we respectively show ZF-UDS-7329 with the Milky Way-like Kroupa and bottom-heavy IMFs. The vertical dashed line indicates the lower limit achieved in the ATLAS$^{3\text{D}}$ sample for z$\sim$0 measurements.}
\label{fig:time-scale}
\end{figure}

One way to obtain the formation time scales of galaxies is by considering their SFH, with t$_{50}$ defined as the time it took to form 50\% of the stellar mass of the galaxy since the first episode of star formation. Alternatively, the typical formation time scales of quiescent galaxies can be approximated by using the abundance of [Mg/Fe] through the expression [Mg/Fe]$\sim1/5 - 1/6$~log$\Delta$t$_{[\text{Mg}/\text{Fe}]}$ \citep[e.g.][]{Thomas2005}. In Figure~\ref{fig:time-scale}, we show the comparison between the chemical formation time scale inferred this way and t$_{50}$ for the ATLAS$^{3\text{D}}$ sample. The colours of the symbols correspond to the mass of the ATLAS$^{3\text{D}}$ galaxies (M$_{\text{JAM}}$). One can see the trend that more massive galaxies have faster formation time scales. The mismatch between the two ways to quantify the formation time scale is evident, with t$_{50}$ being systematically higher and unable to probe time scales shorter than $\sim$0.6~Gyr (vertical dashed line) due to the lack of age sensitivity.

Conversely, the detailed analyses of the high-z galaxy ZF-UDS-7329 shows a notable agreement between the chemical and SFH-based formation time scales, as shown in Fig.~\ref{fig:time-scale}. The SFH-based formation time-scale values were obtained using the Milky Way-like Kroupa and bottom-heavy IMFs, respectively in orange and blue in the figure. The resolution of the NIRSpec-PRISM spectrum is too low to reliably measure [Mg/Fe]. The chemical formation time scales were obtained by adopting the abundance reported in \citet{Carnall2024} of [Mg/Fe]=0.42$_{-0.17}^{+0.19}$, which is based on the spectra taken with the three NIRSpec medium-resolution gratings. From this we obtained a $\Delta$t$_{[\text{Mg}/\text{Fe}]}$=0.26$^{+0.30}_{-0.22}$. This is in agreement with the SFH shown in Fig.~\ref{fig:ppxf_fit}, where half of its stellar mass was formed within 0.25 Gyr (i.e. the size of the first age bin). This agreement between the chemistry and SFH-based time scales only becomes possible thanks to the quiescent yet young stellar populations of ZF-UDS-7329.

\section{Conclusion} \label{sec:conclusion}

We have analysed the star formation history and stellar population properties of the high-redshift massive galaxy ZF-UDS-7329 by employing the full spectral fitting code pPXF and the MILES SSPs as a base. We used two sets of models, one with a Milky Way-like IMF and a second with a bottom-heavy IMF of slope $\Gamma_{\text{b}}=3.0$, which is similar to the one measured on the massive relic galaxy NGC~1277. We find that in both cases, the majority of the galaxy's mass forms very rapidly (less than $0.25$~Gyr) and early in the history of the Universe, with negligible star formation afterwards. We applied a passive ageing process to the galaxy's spectrum and compared it to NGC~1277. We also compared the mass growth of ZF-UDS-7329 with a sample of local quiescent massive galaxies, demonstrating its remarkably fast formation. Finally, we compared the chemical and SFH-based formation time scales of ZF-UDS-7329 and the sample of local massive galaxies, which showed good agreement between them.

The early star formation of ZF-UDS-7329, compatible with the age of the Universe, in a very fast episode and subsequent quenching, its high mass, and the similarities between its aged spectrum and visual resemblance with NGC~1277 all indicate that ZF-UDS-7329 is a high-z precursor of massive relic galaxies or the core of massive ellipticals. The increased time resolution afforded by studying younger stellar populations in high-z galaxies allowed us to probe shorter formation time scales than is currently possible by studying z=0 quiescent galaxies. With this, we achieved a better agreement between the chemical and SFH-based time scales.

Further observations taking advantage of the higher spectral resolution of the other instruments aboard JWST are already on going and will be able to provide detailed abundance patterns and stellar kinematics and will probe the IMF at high redshift. This will allow these characteristics to be traced from very early in the history of the Universe to the present day and results from stellar population modelling to be compared with the currently accepted cosmological model constraining the ages and masses of galaxies at early epochs.

\begin{acknowledgement}

We thank the referee for the careful reading of this letter and the useful comments that helped improve the manuscript. We thank E. Glazebrook for sharing the data that was the basis of this work and R. McDermid for the ATLAS$^{3\text{D}}$ data. EAH, IMN and AV acknowledge support from the PID2022-140869NB-I00 grant from the Spanish Ministry of Science and Innovation. MHC and PIN acknowledge financial support from the State Research Agency of the Spanish Ministry of Science and Innovation (AEI-MCINN) under the grants “Galaxy Evolution with Artificial Intelligence” with reference PGC2018-100852-A-I00 and “BASALT” with reference PID2021-126838NB-I00. JPVB received a fellowship from the "la Caixa" Foundation (ID 100010434), the fellowship code is LCF/BQ/DI23/11990084. MM acknowledges support from the grant RYC2022-036949-I financed by the MICIU/AEI/10.13039/501100011033 and by ESF+.
\textit{Software:} We acknowledge the use of the python packages \textsc{numpy} \citep{Numpy}, \textsc{matplotlib} \citep{Matplotlib} and \textsc{astropy} \citep{Astropy}. 

\end{acknowledgement}

\bibliographystyle{aa}  

\end{document}